# Maturity of the Internet of Things Research Field: Or Why Choose Rigorous Keywords

**Marta Vos**
School of Business and IT
Whitireia International
Porirua, New Zealand
Email: marta.vos@whitireia.ac.nz

## Abstract

Periodically researchers examine the maturity of their research fields against a set of maturity criteria including the types of theories and methods used within the field. Keyword metadata was used to locate relevant research articles, which were manually read, summarised, and compared against field maturity criteria. This study used data extracted from three academic databases to examine the maturity of the Internet of Things (IoT) research field using keyword metadata analysis alone. The metadata analysis was proposed to reduce analysis time, and allowed for testing of a much larger sample of articles. Findings indicated that the IoT research field was immature, with experimental methods dominating research outputs. Difficulties with keyword metadata including synonyms, abbreviations, poor accuracy, and method and theory not being included in metadata, made this research challenging. To relieve these problems in the future a keyword schema is suggested. This should also assist researchers in locating relevant literature.

## 1 INTRODUCTION

As a field of research develops it is common to periodically examine and summarise it through literature surveys and maturity studies. This examination helps researchers determine how the field is maturing, to identify any developing theory and methodologies, and to uncover any research gaps. However, as the body of literature increases manual examination of a broad range of literature becomes increasingly difficult because of the sometimes overwhelming quantity of literature available to the researcher. When the quantity of literature available is too great the researcher must cut down the amount of literature being reviewed through applying strict criteria such as limiting the survey to certain journals, full text articles, imposing date or topic ranges, and discarding articles based on their abstracts in order to be able to carry out a manual analysis (Chu, 2015; Michelberger, Andris, Girit, & Mutschler, 2013). This manual analysis generally consists of reading the entire article and coding it for specific features depending on the focus of the study being undertaken. Maturity studies usually take a similar path with researchers examining the literature within the topic field in order to determine how mature research is within that field. This assessment is made based on a variety of criteria such as the types and combinations of research methods used, the number of authors and papers published, and the development of new theory (von Krogh, Rossi-Lamastra, & Haefliger, 2012).

The majority if not all academics utilise academic library databases to assist them in their literature searches. These databases store a variety of metadata information such as abstract, keywords and article title, along with the actual content of the academic journal articles, and this information can be searched in a variety of different ways (Brand, Daly, & Meyers, 2003). This study was undertaken in an attempt to determine if analysis of this metadata alone could be useful in summarising a field of research and determining its maturity, without requiring any manual intervention through a comprehensive reading of the articles identified by the researcher. As the development of the Internet of Things (IoT) is rapidly gaining momentum and generating a great deal of interest, and because it seemed appropriate to use a metadata study to examine a research field which involves the generation of large data sets, the IoT was chosen as the unit of analysis for this research.

Although there is no definitive definition of the Internet of Things (IoT), it is generally considered to be a network of digitalised IT artefacts (things) connected to each other, the internet, and humans, generally through the use of RFID technology (Atzori, Iera, & Morabito, 2010; Vos, 2014). In recent years the IoT has attracted a large amount of research effort with a mixture of technical developments, and attempts to understand the social and management implications of what is a rapidly developing potentially disruptive technology, the boundaries of which have yet to be fully explored (Atzori et al., 2010; Li et al., 2014). Therefore, this study was carried out initially in an attempt to use metadata analysis to understand the maturity of the IoT field, and to gain some understanding of the types of methods that were being used by researchers to study the field through metadata analysis.
The research questions are therefore derived as:

RQ1 – How mature is the IoT research field?

RQ2 – How effective is metadata analysis in determining the maturity of the IoT research field?

# 2 LITERATURE REVIEW

As the storage capacity of information systems increases the problem of storing ever increasing amounts of information decreases. However, difficulties with retrieving the right information are increasing concurrent with the volume of information available (Rorissa & Yuan, 2012). These difficulties are becoming apparent in the analysis of research literature with researchers such as Palvia et al. (2004) listing amongst the limitations of their analysis the inability to query a large number of journals in order to determine the trends apparent in MIS research. This problem occurred because of difficulties associated with handling a data analysis where each of the articles identified by their search had to be manually read and coded.

Williams et al. (2009) in a survey of IT adoption and diffusion research, queried article titles in order to locate their research literature. The resultant 10,000 plus articles, located despite the search requiring the presence of three key words in the title, were narrowed down by limiting the search to just the IS/IT field, and then further through journal based selection to a total of 345 articles. They then conducted a manual analysis of abstract metadata to determine research paradigm, data issues and units of analysis for each selected article. More comprehensive literature surveys such as that of Irani, Gunasekaran and Dwivedi (2010) who surveyed 666 articles in the RFID research field, are carried out in order to gain an overall insight into a particular topic area, and to identify research trends, theories, and gaps in the literature. Whitmore, Agarwal and Xu (2014) manually examined 127 articles in the IoT field, finding that the literature was largely focused on technology, while the discussion of management and governance was neglected. They developed a literature framework composing of technology, applications, challenges, business models, future directions and overviews/survey categories. These types of summaries, with analysis of theory and method can also give researchers an insight into the maturity of the particular field of study.

## 2.1 Research Field Maturity Studies

Cheon, Grover and Sabherwal (1993) argue that the maturity of a field can be assessed based on what the field says about itself through its body of research literature. They suggest three measures to assess the maturity of a field of research including the range of variables researched (and construction of paradigms), a range of methodologies being used, and hypothesis testing (rather than description) being the features of a mature field of research.

In considering field research Edmonson and McManus (2006) determined three levels of research maturity. Nascent maturity, characterised by mainly qualitative study with little research available discussing the field being investigated. Intermediate maturity, characterised by more quantitative study with some constructs, provisional theory, and some measures developed. Finally mature study characterised by mixed methods, an extensive literature and tested constructs and measures. Similarly, Krogh, Rossi-Lamastra and Haefliger (2012), in the case of phenomenon based research, recognise three phases of maturity, the embryonic, growth and mature. They recognised the embryonic phase as being characterised by a few researchers focusing on the phenomenon, with un-coordinated, and duplicated efforts. The growth phase sees the phenomenon solidifying with theory developing. In the mature phase the characteristics of the phenomenon emerge along with a variety of research methods and approaches. In the field of library and information sciences Chu (2015) found that the field was maturing as it moved towards a wider variety of research methods, away from the traditional historical and questionnaire based analysis.

The similarities between these different ways of measuring maturity within a research field suggests a generalised set of maturity criteria would be appropriate to test against the IoT research field. Thus the maturity criteria used will be as follows:

1. Immature Phase: lower range of topics and methodologies, with a few researchers focusing on the area.

2. Growth Phase: a range of methodologies with theory developing.
3. Mature Phase: quantitative hypothesis testing with a wide variety of research methods and approaches.

### 2.2 The Use of Keyword Metadata

Generally, when assessing the maturity of a research field, researchers assess the existing literature against maturity criteria, assuming that what researchers say about themselves through their articles and associated metadata is the most accurate way to conduct this kind of study (Romano Jr, 2001). Rorissa and Yuan (2012) also believe that author assigned keywords reveal the areas that researchers are focusing on at any particular time, giving an accurate picture of the progress of current research. Kevork and Vrechopoulos (2009) found that author assigned keywords give a good indication of the subject of the article and are unbiased since they don't rely on a researchers interpretation of the text of an article. Similarly, Romano and Fjermestad (2001) found that using keywords for analysis allowed the context of the article to be retained and produced better search results because of this. In the field of tourism, Wu et al. (2012) undertook an analysis of tourism literature in order to determine core research areas. They recommended the use of author selected keyword analysis believing that these keywords helped "capture the theme or essence of a piece of research" (Wu et al., 2012, p. 359). They also believed that such keywords may give insight into the location, methodology, theory, and findings of a piece of research.

However, the use of keyword analysis also has its detractors. LaBrie (2014) lists a number of issues with the use of keywords including infrequency of use, the use of synonyms, ambiguous terms, keyword phrases which are too long, or keywords which include inappropriate special characters. Keyword classification schemes are used by many databases, although some have dropped them completely as being problematic, including MIS Quarterly (LaBrie, 2014). Wu et al. (2012) also highlighted problems with keyword searching including varying capitalisation and plural usage, abbreviations and the use of gerunds, as well as the use of synonyms by authors. Masuchika (2014) focused on the use and creation of synonyms by researchers, highlighting the difficulties this creates for those wishing to search for and identify relevant research articles. Despite this issue, he did not call for the creation of a limited keyword vocabulary believing that this would inhibit the creation of new terms, or varying terms when they are legitimately needed.

Although there are a number of studies assessing field maturity partially through the use of keywords, such as that of Romano (2001) none could be located that used keyword metadata alone to assess maturity. Therefore, this study attempts to assess the maturity of the IoT field against the maturity criteria listed in Section 2.1 above, using only keywords and other metadata assigned to IoT research articles.

## 3 METHOD

According to the Dublin Core Metadata Initiative (DCMI) metadata includes among other things the title, authors, subject/s (keywords), and description (abstract) of an article (Brand et al., 2003). In many academic databases an articles metadata includes the title, authors, journal, volume, issue and pages, in fact all non "full text" parts of the research article.

In order to obtain data for this study a data extract was obtained from three academic databases, based on "internet of things" as an author assigned keyword. The IEEE, ProQuest and Scopus databases were chosen in order to represent a range of social and technical approaches to studying the internet of things. They were also chosen because of the need to be able to extract metadata to a spreadsheet, an option not offered by all academic databases. For all three databases the search was limited to academic peer reviewed articles or conference papers. Data extracts included author assigned keywords, standard keywords, article title, abstract, and publication information. This information makes up all but full text metadata for most article database searches.

Once the data was extracted to spreadsheet duplicated articles were removed through identification of journals held in more than one database. Keywords were then separated and the keyword data was cleansed by removing extra spaces at the beginning of words, and converting all words to lower case. Random special characters such as * or # were removed but brackets and "-" were retained as these can form reasonably common parts of keywords.

Keywords were analysed by sorting into broad areas based on the focus of the keyword. In order to identify the methods and theories being used in the field keywords were initially searched. However, the lack of keywords referring to method or theory led to a broader manual re-search of the article metadata using a list of methods and theories common in IT and IS research.

# 4 FINDINGS

In total 8,168 articles were identified across the three databases as summarised in Table 1 below, using "internet of things" as an author assigned metadata keyword. These articles all together had 37,024 keywords in varying combinations with 12,019 individual unique keywords.

| Database | Total Articles | Total Keywords |
|---|---|---|
| IEEE | 1,670 | 8,763 |
| PQ | 1,269 | 7,139 |
| Scopus | 5,229 | 21,122 |
| Totals | 8,168 | 37,024 |

*Table 1: Total number of articles and keywords by database*

A visualisation of the raw keyword data is presented in Figure 1 below. This visualisation demonstrates some of the difficulties with handling the raw keyword data in that the conformation of the keywords used, including capitalisations and abbreviations varying substantially between authors.

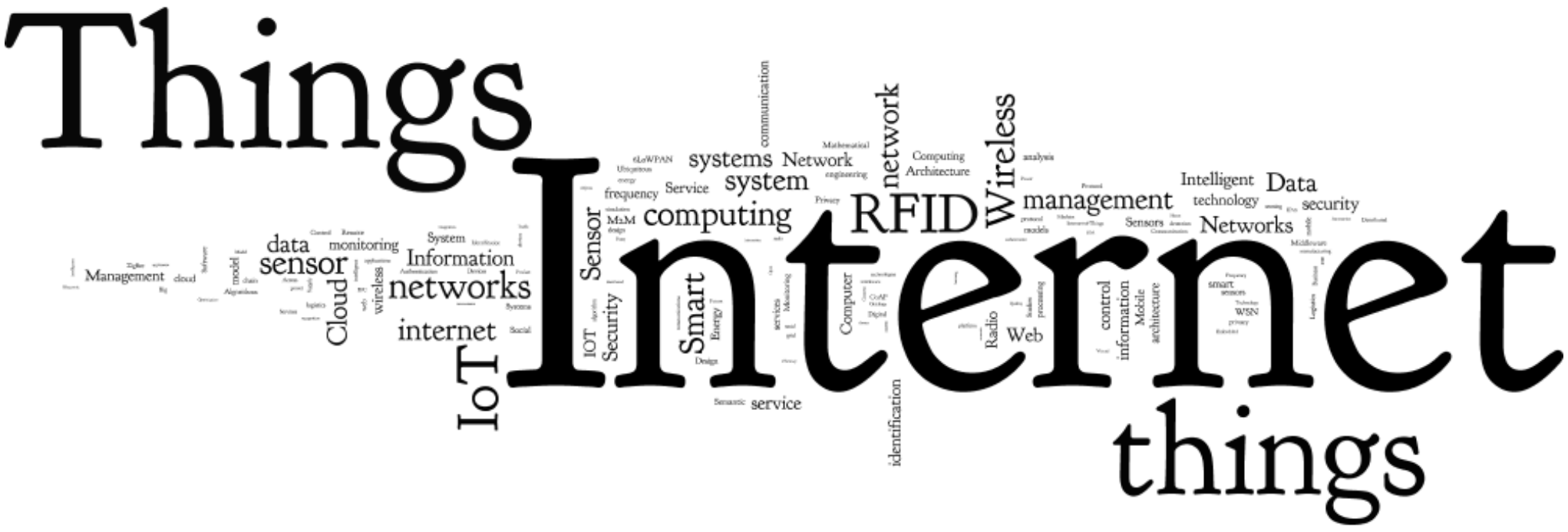

*Figure 1: Visualisation of raw keyword metadata*

There were also numerous synonyms found for the same keyword, most commonly with the keyword "internet of things" itself. Even when the data was cleaned of capitalisations and some special characters, there were a total of 23 different synonyms and abbreviations used for the internet of things, the most common being presented in Table 2 below.

| Variation on Internet of Things | |
|---|---|
| internet of things | 3,862 |
| internet of things (iot) | 482 |
| iot | 405 |
| internet-of-things | 92 |
| interenet of things - and another keyword combined | 61 |
| internet of thing | 29 |
| internet of things(iot) | 28 |
| internet of things technology | 25 |
| internet-of-things (iot) | 24 |
| iot (internet of things) | 14 |
| internet of thing (iot) | 8 |
| internet of things (iots) | 8 |
| internet of things - mis-spellings and odd characters | 12 |
| **Total** | 5,050 |

*Table 2: Variations on "internet of things" keyword*

The number of articles published each year, along with the number of peer-reviewed journals and conferences publishing them, is continuing to increase dramatically as demonstrated in figure 2 following:

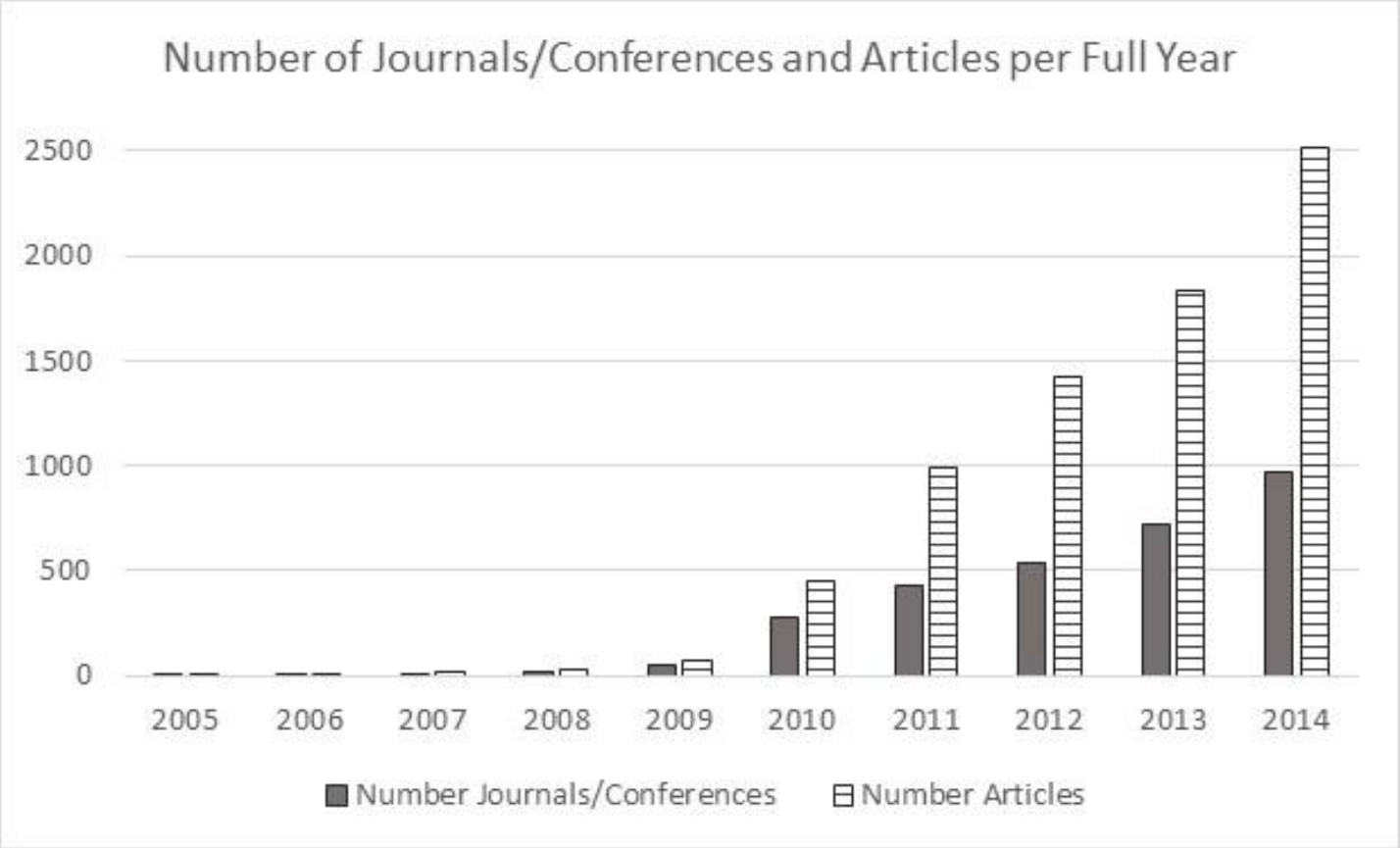

*Figure 2: Graph demonstrating number of articles published and journals/conferences per year*

A total of 2,749 individual publications and conferences published 8,168 articles related to the internet of things, many only publishing once. As can be seen both the number of articles/conference papers, and the number of journals/conferences is increasing per year. The top 5 journals/conferences are presented below in Table 3 below:

| Journal | Number Articles |
|---|---|
| Lecture Notes in Computer Science | 298 |
| Advanced Materials Research | 286 |
| Communications in Computer and Information Science | 99 |
| Lecture Notes in Electrical Engineering | 81 |
| Jisuanji Xuebao/Chinese Journal of Computers | 57 |

*Table 3: Journals/conferences publishing most internet of things articles*

As can be seen from Table 4 below, the most prolific author name has a total of 82 articles over a 5 year period. However, the lack of a first name in the databases queried means that there is no way of knowing how many authors are included within “Zhang Y”. It is unlikely that these top five authors represent five individuals.

| Year | 2010 | 2011 | 2012 | 2013 | 2014 | Total |
|---|---|---|---|---|---|---|
| Zhang Y. | 8 | 11 | 25 | 19 | 19 | 82 |
| Wang Y. | 7 | 11 | 17 | 16 | 16 | 67 |
| Wang J. | | 14 | 19 | 14 | 12 | 59 |
| Liu Y. | 10 | | 13 | 17 | 17 | 57 |
| Wang X. | | 7 | 12 | 18 | 14 | 51 |

*Table 4: Author names per year*

The keyword data was sorted into category groups in order to determine the overall research emphasis. Because 12,019 individual keywords were identified, only those keywords used more than 5 times were included in this analysis. The following Table 5 lists the top 5 keywords in each category. The technology category has additional entries because of the dominance of this category. The categorisation scheme for the internet of things developed by Whitmore, Agarwal and Xu (2014) was not used because of the amount of technological metadata uncovered by this study, and the relative lack of any metadata in respect of business models/future directions or overviews. Instead categories covering context, data, research, and social metadata appeared more logical.

| **Applications** | | **Research** | | **Technology** | |
|---|---|---|---|---|---|
| smart cities | 77 | ontology | 76 | internet of things related | 5,624 |
| smart city | 75 | mathematical analysis | 45 | network related | 2,710 |
| mhealth | 19 | game theory | 16 | rfid related | 1,202 |
| ehealth | 17 | ontologies | 10 | internet | 657 |
| intelligent monitoring | 13 | research | 10 | cloud | 464 |
| **Total** | **920** | **Total** | **255** | security | 282 |
| **Context** | | **Social** | | standards | 276 |
| china | 37 | privacy | 150 | sensors | 251 |
| contiki | 15 | social networks | 59 | architecture | 138 |
| evolution | 15 | future internet | 57 | 6lowpan | 129 |
| coal mines | 15 | trust | 31 | algorithms | 126 |
| foods | 13 | e-health | 29 | zigbee | 121 |
| digital city | 10 | environmental monitoring | 28 | big data | 108 |
| **Total** | **301** | **Total** | **789** | wireless networks | 106 |
| **Data** | | **Management** | | monitoring | 102 |
| computation | 42 | management | 93 | smart grid | 102 |
| data mining | 29 | logistics | 79 | devices | 97 |
| data processing | 29 | interoperability | 66 | coap | 96 |
| data management | 25 | standards | 57 | ubiquitous computing | 96 |
| linked data | 25 | information systems | 53 | authentication | 91 |
| **Total** | **1,151** | **Total** | **1,536** | **Total** | **17,176** |

*Table5: Grouped keywords*

The keyword metadata also presented difficulties when assessing field maturity. The majority of articles did not have the method included in keywords, requiring a second manual search through the metadata using a list of methods and theories common in IS and IT, in order to determine the methods and theories used by authors. This data has been summarised according to the type of method or theory mentioned, and is presented in Table 6 following. The number of times a method or theory was mentioned in keyword data compared to the number of metadata mentions located through manual searching in other metadata fields (primarily in abstract but occasionally in title) are included separately, along with an indication of the number of keyword variants used for the keyword.

| | Total No. Metadata Instances | Total No. Keyword Instances | Keyword Variants | | Total No. Metadata Instances | Total No. Keyword Instances | Keyword Variants |
|---|---|---|---|---|---|---|---|
| **Quantitative Methods** | | | | **Experimental/Mathematical** | | | |
| Quantitative | 78 | 3 | 2 | Experiment | 1054 | 32 | 19 |
| Statistical analysis | 17 | 2 | | Mathematical model | 119 | 119 | |
| Questionnaire | 8 | 0 | | Neural network | 55 | 35 | 13 |
| Cluster analysis | 5 | 3 | | Game theory | 28 | 20 | 3 |
| Structural equation modelling SEM | 3 | 1 | | NS2 Simulation | 9 | 0 | |
| Mixed method | 1 | 0 | | Regression analysis | 5 | 5 | |
| **Total** | **112** | **9** | | Factor analysis | 3 | 1 | |
| **Qualitative Methods** | | | | Probability theory | 3 | 3 | |
| Qualitative | 40 | 2 | 2 | Chaos theory | 1 | 1 | |
| Case study | 198 | 3 | | Input-output analysis | 1 | 0 | |
| Conceptual model | 26 | 2 | | **Total** | **1278** | **216** | |
| Interview | 21 | 1 | | **Theoretical** | | | |
| Content analysis | 5 | 1 | | TAM | 8 | 6 | 3 |
| Delphi | 5 | 3 | 2 | Actor network theory | 2 | 0 | |
| Action research | 4 | 0 | | DOI | 2 | 1 | |
| Literature survey | 2 | 0 | | Task-technology fit | 1 | 0 | |
| Ethnography | 1 | 1 | | TOE | 1 | 1 | |
| Grounded theory | 0 | 0 | | UTAUT | 1 | 0 | |
| **Total** | **302** | **13** | | **Total** | **15** | **8** | |

*Table 4: Method and theory metadata instances*

Nb. Each instance represents one article – rather than a total number of times the word was used in the data. Where there are abbreviations listed, both the abbreviation and full text were searched.

# 5 DISCUSSION

This study used a data extract of author assigned keywords and other article metadata to examine the research field of the internet of things, in order to determine its maturity against the following maturity criteria:

1. Immature Phase: lower range of topics and methodologies, with a few researchers focusing on the area.
2. Growth Phase: a range of methodologies with theory developing.
3. Mature Phase: quantitative hypothesis testing with a wide variety of research methods and approaches.

The findings clearly show that the field of the internet of things is still in the immature phase. The number of articles published, and numbers of journals and conferences publishing them are still increasing rapidly. However, it was not possible with the data available to comment on the number of authors publishing. The number of topics being examined is still relatively low, with a heavy emphasis on the technical and experimental, while the management, context and social implications of the internet of things are relatively neglected.

The most telling evidence of the maturity of the field comes from an analysis of methodology and theory. Where a possible method could be presumed, the experimental method with 1,278 of 8,168 articles mentioning the word "experiment" in the article metadata was by far the most common. In terms of either qualitative or quantitative methods only 112 and 302 articles respectively contained metadata mentions that could be considered to fall in these areas. The most used non-experimental method was case study, fitting well with the contention of Yin (2009) that case study is most commonly used where little is known about a research field. Theory was even less mentioned, with specific theory mentions in only 15 articles, and a number of theories, such as institutional theory, IS success, social shaping of technology and adaptive systems theory, not mentioned at all. This evidence points towards an immature field with little theory basis, or study of the organisational effects of the technology.

However, analysing the data extract was more difficult than initially expected. It is apparent from the data summarised in Table 6 that many authors did not mention theory or method in keyword metadata. It is also probable that theory and method were not mentioned in other metadata (including abstract), although the methodology of this study did not allow time for the reading of a sample of articles in order to determine if this was the case. Wu et al. (2012) also observed that not all articles they examined included keywords related to the theory used by the authors. It was also not possible to assess whether or not the field was using multi or mixed method research given the research design as it would have been necessary to read the research articles in more depth. “Mixed method” did not appear in the research metadata.

It was not the intent of this research to read the articles in order to confirm findings. Instead the intent was to demonstrate the utility if any, of deriving an understanding of a field and an assessment of its maturity, based on an analysis of research article metadata. Author assigned keywords were initially queried as these form the basis of many literature searches, as discussed in the literature review above. However, the poor quality of the keyword metadata with multiple synonyms, a lack of descriptive rigor and the presence of sometimes random symbols and words, caused problems in both selection and analysis of the articles. Therefore a schema for those considering keyword selection is suggested as follows:

1. Use standard descriptors for the field
2. Do not use special characters
3. Consider including keywords representing:
    a. Key technological actor/s
    b. Key unit/s of analysis
    c. Key context/s
    d. Key method/s
    e. Key theory/s
    f. Key finding/s

As found by Barki, Rivard and Talbot (1993) the lack of a common vocabulary makes searching for research information more difficult. This finding still holds true in todays age of internet metadata search. While it is possible to recall a great deal of information in a search, finding the right information is a different matter. Some thought applied by authors to the selection of keyword metadata would add considerably to the ease of both finding research articles, and to analysing research fields.

# 6 CONCLUSION

As pointed out by a number of authors of meta-analysis papers, a limitation of many of their studies is the time taken to manually process each research article determining the topics it covers, methods used, and theoretical basis for research. Because of the time consuming nature of this analysis, a large number of articles are generally excluded for one reason or another in order to accommodate the manual analysis.

It was hoped that the ability to extract and analyse a large metadata set of peer reviewed journal articles and conference papers would lead to shortening the time taken to analyse the internet of things research field, and to present a more efficient way to analyse its maturity. Unfortunately the low quality of the metadata analysed detracted from some time savings that might be made through using metadata analysis. Although topic analysis is possible, and takes considerably less time than the semester long manual analysis of Palvia et al (2004), methodology and theory analysis was impaired by the number of authors that failed to mention either their methodology or theory within the metadata of their articles. The inclusion of a much larger data set did allow for a comprehensive overview of the topics published within the research field of the internet of things, and despite problems with the metadata which mean this type of analysis is likely less accurate than manual analysis, it is still apparent that the IoT research field is in the immature phase.

There are a number of limitations present in this study. Firstly, it was not possible from the metadata to determine the nature of many of the more common methods. For example experiments and surveys could have a wide range of meanings and it was impossible without a comprehensive reading of the articles to determine exactly what kind of experiments and surveys were being conducted. It is also likely that those papers that were published in conferences then subsequent journals would have double entries and no reasonable way could be determined to detect this as often content and even authors can change with these articles.

This study contributes to research by highlighting the relatively immature state of internet of things based research, finding that there is still a great deal to be done to understand the organisational and social impact of this phenomenon. Further, this study recommends a schema for authors to consider when selecting keywords in order to make this kind of analysis easier, and more importantly, to make it easier for others to locate their articles through the use of good keyword selection.

## Acknowledgements

I would like to thank Amanda Crossham and Dr Nicole Gaston for assisting me in finding references from the Information Studies discipline.